\documentclass[12pt,preprint,showpacs,superscriptaddress,pre]{revtex4-1}
\usepackage{amsmath}
\usepackage{amssymb}
\usepackage[dvips]{graphicx}
\usepackage[svgnames]{xcolor}
\usepackage[colorlinks=true,citecolor=blue,linkcolor=blue,citebordercolor=white,linkbordercolor=white]{hyperref}
\usepackage{enumerate}
\usepackage{natbib}

\def\eq#1{{Eq. (\ref{#1})}}
\def\sign{{\rm sign}}

\begin{document}

\title{Diffusion of a particle in the Gaussian random energy  landscape: Einstein relation and analytical properties of average velocity and diffusivity as functions of driving force}

\author{S.V. Novikov}
\email{novikov@elchem.ac.ru}
\affiliation{A.N. Frumkin Institute of
Physical Chemistry and Electrochemistry, Leninsky prosp. 31,
119071 Moscow, Russia}
\affiliation{National Research University Higher School of Economics, Myasnitskaya Ulitsa 20, Moscow 101000, Russia}



\begin{abstract}
We demonstrate that the Einstein relation for the diffusion of a particle in the random energy landscape with the Gaussian density of states is an exclusive 1D property and does not hold in higher dimensions. We also consider the analytical properties of the particle velocity and diffusivity for the limit of weak driving force and establish connection between these properties and dimensionality and spatial correlation of the random energy landscape.
\end{abstract}

\maketitle


\section{Introduction}

Fundamental feature of a simple diffusion process is the validity of the Einstein relation (ER) between the diffusivity $D$ and drift mobility $\mu$. A particular but very important example of the diffusive transport is the motion of charge carriers in amorphous semiconductors under the action of the applied electric field $E$, where the ER takes form
\begin{equation}
D= \frac{kT}{e}\mu.
\label{ER}
\end{equation}
Relation (\ref{ER}), apart from the clear fundamental importance, serves as a very useful tool for the estimation of $D$ in many materials demonstrating hopping charge transport. Indeed, the mobility could be rather easily measured in experiments, e.g. by the time-of-flight technique, while the direct measurement of $D$ is much more difficult. At the same time, in many materials the mobility depends on $E$ and the simple Einstein relation (\ref{ER}) is not valid. It was found that for the case of the 1D transport in disordered materials with the Gaussian density of states (DOS) the properly modified Einstein relation is valid \cite{Parris:5295,Parris:2803}
\begin{equation}
D= \frac{kT}{e}\frac{\partial V}{\partial E},
\label{mER}
\end{equation}
where $V$ is the average carrier velocity. This very relation may be rewritten in a more beautiful form. Indeed, if we let the magnitude of disorder goes to zero while keeping all other relevant parameters the same, then for the resulting system  the simple Einstein relation $D_0=kT\mu_0/e$ is certainly valid (here the corresponding diffusivity and average velocity are $D_0$ and $v=\mu_0 E$), so \eq{mER} is equivalent to
\begin{equation}
\frac{D}{D_0}= \frac{\partial V}{\partial v}.
\label{mER-2}
\end{equation}
In this form the modified Einstein relation (mER) contains no parameters such as $e$, $T$, etc. In future we will use this very form of the mER. A natural question is whether \eq{mER-2} could be extended to the multidimensional case. In this paper we are going to demonstrate that the mER is strictly the 1D relation which could not be extended to higher dimensions.

We consider the continuous model of the carrier diffusion in the random energy landscape $U(\vec{x})$. It provides a proper description of the long time behavior of the hopping charge carrier transport. In fact, in the strict sense nether ER nor mER is valid for the lattice model of the hopping transport. Indeed, let us consider the simplest model of the hopping to the nearest neighbors only for the 1D chain without disorder and for the Miller-Abraham hopping rate \cite{Miller:745}. In this case a simple calculation gives for the velocity $v$ and diffusivity $D$
\begin{eqnarray}
\label{vD}
  v &=& \nu_0 a \hskip2pt\sign(E)\left(1-e^{-|\lambda|}\right), \\
  D &=& \frac{1}{2}a^2\nu_0\left(1+e^{-|\lambda|}\right).
\end{eqnarray}
here $\nu_0$ is the scale of the hopping rate, $a$ is the lattice scale, $\lambda=eEa/kT$. Both Eqs. (\ref{ER}) and (\ref{mER}) are invalid for this model, apart from the limit $\lambda \ll 1$. This phenomenon is not a specific property of the Miller-Abraham hopping rate, because the use of an arbitrary hopping rate leads to the substitution $\nu_0\to\nu_0 f(|E|)$, with some function $f(|E|)$ going to a constant at $E\to 0$. Again, Eq. (\ref{ER}) or (\ref{mER}) hold only in the limit $\lambda \ll 1$. We may conclude that the ER or mER are not valid for the lattice hopping models even in the ideal case of absolutely ordered 1D lattice. For this reason we limit our consideration to  the continuous diffusion model.

To avoid a possible confusion we mention here another generalization of the Einstein relation, typically called the generalized ER (gER), which is specifically tailored for the charge transport in the case of not very low charge density  \cite{Hope:2377,Roichman:1948,Nguyen:1998,Tessler:200,Wetzelaer:66605}. The gER for the Gaussian DOS has the form
\begin{equation}
D= g(n,T)\frac{kT}{e} \mu,
\label{gER}
\end{equation}
where the enhancement factor $g$ depends on the carrier density $n$ and $T$. In contrast to \eq{gER}, the relation (\ref{mER}) is valid  for $n\to 0$ and arbitrary $E$, while the relation (\ref{gER}) is valid only in the case of field-independent $\mu$ which typically implies $E\to 0$.

In addition we are going to consider the dependence of $D$ and $V$ on $v$ for $v\to 0$ (or, equivalently, for $E\to 0$). In recent papers by Nenashev \textit{et al.} a striking difference was found for the dependence of the hopping carrier velocity and diffusivity for the well known Gaussian Disorder Model (GDM) on $v$ for different dimensionality of space  \cite{Nenashev:115203,Nenashev:115204}. In the first paper the exact solution of the lattice 1D version of GDM has been extensively studied and it was found that $V$ and $D$ are non-analytical functions of $v$
\begin{eqnarray}
V=A_v(T)v+B_v(T)|v|v+...,\\
D=A_D(T)+B_D(T)|v|+...\nonumber
\label{vD}
\end{eqnarray}
instead of the expected behavior
\begin{eqnarray}
\label{vD_exp}
V=A_v(T)v+B_v(T)v^3+...,\\
D=A_D(T)+B_D(T)v^2+...\nonumber
\end{eqnarray}
At the same time, for the 2D and 3D cases the careful numerical simulations and approximate analytic consideration suggest that \eq{vD_exp} provides the proper description of the dependence of $V$ and $D$ on $v$ \cite{Nenashev:115204}. The reason for the exceptional behavior in 1D case is not clear. We are going to clarify the situation and try to answer the question whether the 1D case is indeed exceptional. In addition, we are going to study how the analytical properties of $V(v)$ and $D(v)$ depends on the correlation properties of the random energy landscape. There is a natural reason to expect such connection because in 1D case it is well known that the functional dependence of $V$ and $D$ on $v$ is directly governed by the correlation function $C(x)=\left<U(x)U(0)\right>/\sigma^2$ \cite{Dunlap:542} and computer simulation supports that connection in 3D case, too \cite{Novikov:2584,Novikov:954} (here $\sigma^2=\left<U^2\right>$ is the variance of the disorder and we define the correlation function in such a way that $C(0)=1$). From the general point of view the GDM is just one particular case of the correlated disorder where for site energies $U_i$ the binary correlation function is zero for different sites: $\left<U_i U_j\right>\propto\delta_{ij}$.

A major limitation of our approach is the use of perturbation theory (PT). Yet, we will see that the PT approach for the 1D case gives the proper functional dependence of $V$ and $D$ on $v$, and the corresponding perturbative coefficients $A^{pt}_{v,D}$ and $B^{pt}_{v,D}$ could be obtained by the expansion of the exact coefficients in series in the disorder strength parameter $g=(\sigma/kT)^2$.

At the same time, the result of Ref. \onlinecite{Nenashev:115203} provides a reliable anchor point for the comparison of our results with the exact solution of the particular model. Indeed, the general structure of the functional dependence $V(v)$ and $D(v)$ for the 1D GDM in the limit case $v\to 0$ does nor depend on the disorder strength parameter $g$. We will see that this is a general phenomenon for 1D hopping transport for any type of the correlation function. There is a general agreement that the effect of disorder on the charge carrier transport is the most prominent in the 1D case because in this case the path is predetermined and carrier inevitably has to move across all fluctuations of the random energy landscape. As a result, for all transport parameters ($V$, $D$, etc.) the effect of the strength of disorder becomes weaker when the dimensionality of space $d$ becomes higher. For example, the renormalization group analysis gives the leading asymptotics for $\mu$ and $D$ for $v\to 0$ \cite{Deem:911}
\begin{equation}
\label{D-C-result}
\ln \mu,D \simeq -\frac{1}{d}\left(\frac{\sigma}{kT}\right)^2,
\end{equation}
which agrees well with the exact solution of the 1D case \cite{Dunlap:542,Parris:2803} and computer simulation for 3D case \cite{Novikov:4472,Novikov:954,Novikov:2532}. For this reason we may expect that if the functional form of $V(v)$ and $D(v)$ for $v\to 0$ does not depend on the strength of disorder in 1D case and, hence, the PT approach provides the true functional form of $V(v)$ and $D(v)$ for $v\to 0$, then the same is true for any $d$.

\section{Einstein relation}

Let us consider diffusion of a particle in $d$-dimensional space with the random energy landscape $U(\vec{x})$ having the spatially correlated Gaussian DOS. For the particular realization of $U(\vec{x})$ the particle  Green function $G_U(\vec{x},t)$ obeys the equation
\begin{equation}
\frac{\partial G_U}{\partial t}=D_0\nabla\cdot\left[\nabla G_U+\beta G_U \nabla U\right]-\vec{v}\cdot\nabla G_U,\hskip10pt G_U(\vec{x},0)=\delta(\vec{x}),\hskip10pt \beta=\frac{1}{kT}.
\label{G(x,t)}
\end{equation}
We are going to consider the perturbation theory expansion for the Green function $G(\vec{k},s)=\left<G_U(\vec{k},s)\right>$ averaged over static disorder ($G(\vec{k},s)$ is the Fourier transform of the Green function on $\vec{x}$ and Laplace transform on $t$), the corresponding approach and diagrammatic expansion are briefly described in the Appendix \ref{appA}. We limit our consideration to the stationary state $s=0$ and do not write the argument $s$ anymore. Averaged Green function at $s=0$ is perfectly suitable for the description of the dynamics of the particle in a well established transport regime where the initial relaxation is over and experimentally measured particle velocity and diffusivity do not depend on time anymore. Introducing the self-energy $\Sigma(\vec{k})$ and taking into account the usual representation of $G$
\begin{equation}
G^{-1}(\vec{k})=D_0 k^2+i\vec{v}\cdot \vec{k}-\Sigma(\vec{k}),
\label{G}
\end{equation}
we may calculate the corrections to the effective diffusivity $D=D_0+\delta D$ and average velocity $V=v+\delta V$ as
\begin{equation}
\delta \vec{V}=i\left.\frac{\partial\Sigma}{\partial \vec{k}}\right|_{\vec{k}=0},\hskip10pt
\delta D_{ab}= -\frac{1}{2}\left.\frac{\partial^2\Sigma}{\partial k_{a}\partial k_{b}}\right|_{\vec{k}=0}.
\label{delta_vD0}
\end{equation}
Using the first order correction to self-energy \eq{fSig1e}, we obtain
\begin{equation}
\delta \vec{V}^{(1)}=i\frac{gD_0^2}{(2\pi)^d}\int d\vec{p}C(\vec{p})G_0(-\vec{p})p^2\vec{p}, \hskip10pt g=(\sigma\beta)^2.
\label{delta_v10}
\end{equation}

\begin{equation}
\delta D_{ab}^{(1)}=-\frac{g D_0^2}{(2\pi)^{d}}\int d\vec{p}C(\vec{p})
G^2_0(-\vec{p})\left[\left(D_0p^2+i\vec{v}\cdot\vec{p}\right)p_a p_b-
\frac{i}{2}p^2\left(v_ap_b+v_bp_a\right)\right].
\label{dD10}
\end{equation}
Diffusion tensor for $d > 1$ in the coordinate system where one axis is parallel to $\vec{v}$ is diagonal $\textbf{D}={\rm diag}(D_{||},D_{\bot},...,D_{\bot})$, where $D_{||}$ and $D_{\bot}$ are lateral and transversal diffusion coefficients, correspondingly. Hence,
\begin{eqnarray}
\sum\limits_a \delta D_{aa}=\delta D_{||}+(d-1)\delta D_{\bot}, \\ \sum\limits_{a,b} \delta D_{ab}v_av_b=\delta D_{||}v^2,\nonumber
\label{delta_D_||0}
\end{eqnarray}
and
\begin{equation}
\delta D_{||}^{(1)}=-\frac{g D_0^2}{(2\pi)^{d}}\int d\vec{p}C(\vec{p})
G^2_0(-\vec{p})\left[\left(D_0p^2+i\vec{v}\cdot\vec{p}\right)
\frac{\left(\vec{v}\cdot\vec{p}\right)^2}{v^2}-ip^2
\left(\vec{v}\cdot\vec{p}\right)\right].
\label{delta_D_||_20}
\end{equation}
\begin{equation}
\delta D_{\bot}^{(1)}=-\frac{g D_0^2}{(2\pi)^{d}(d-1)}\int d\vec{p}C(\vec{p})
G^2_0(-\vec{p})\left(D_0p^2+i\vec{v}\cdot\vec{p}\right)
\left(p^2-\frac{(\vec{v}\cdot\vec{p})^2}{v^2}\right).
\label{delta_D_bot0}
\end{equation}

Let us try to extend the mER to the multidimensional case. It is easy to check that the proper extension for the mER is
\begin{equation}
\frac{1}{D_0}\sum_a D_{aa}= \sum_a\frac{\partial V_a}{\partial v_a},
\label{nD-mER-20}
\end{equation}
or, in the proper coordinate system with one axis parallel to $\vec{v}$
\begin{equation}
\frac{1}{D_0}\left(D_{||}+(d-1)D_{\bot}\right) = \frac{\partial V}{\partial v}.
\label{nD-mER-20p}
\end{equation}

This relation is indeed valid for the first order PT, demonstrates a reasonable tensor structure and is the only proper valid extension of the mER which is linear in $D$ and $V$ and does not  explicitly  depend on the effective charge $g$. Unfortunately, this relation does not hold for the second order PT (see Appendix \ref{appB})
\begin{equation}
\sum_a \left(\frac{D_{aa}}{D_0}-\frac{\partial V_a}{\partial v_a}\right)=O(g^2).
\label{nD-mER-20a}
\end{equation}
If the mER is invalid even in the second order PT, then we may safely conclude that the mER is a strict 1D relation having no reasonable extension to the multidimensional case.

Why the 1D mER is valid and what is the difference in the multidimensional case? Diagrammatic approach provides a very clear explanation of this phenomenon. For example, for the 1D case the relation for $\delta\Sigma_2$ simplifies
\begin{eqnarray}
\label{Sigma2_1D}
&\delta\Sigma_2(\vec{k})=k\frac{g^2 D_0^4}{\left(2\pi\right)^{2}}\int\limits_{-\infty}^\infty dp_1 dp_2 C(p_1)C(p_2)\left(p_1 p_2\right)^2 \widetilde{G}_0(k-p_1)\widetilde{G}_0(k-p_1-p_2)\times\\
&\left[\widetilde{G}_0(k-p_1)+\widetilde{G}_0(k-p_2)\right],\nonumber
\end{eqnarray}
here $\widetilde{G}_0(k)=(D_0 k+iv)^{-1}$. Transformation of $kG_0(k)$ to $\widetilde{G}_0(k)$ for every diagram of the PT is the specific feature of the 1D case. Important property of $\widetilde{G}_0(k)$ is
\begin{equation}
\label{deriv_1D}
\frac{\partial \widetilde{G}_0}{\partial k}=-D_0 \widetilde{G}_0^2, \hskip10pt \frac{\partial \widetilde{G}_0}{\partial v}=-i  \widetilde{G}_0^2,
\end{equation}
i.e. these derivatives are proportional to each other and the proportionality coefficient does not contain $k$. Another important property of every diagram is that $k$ (apart from being the common multiplier) is contained here in the arguments of $\widetilde{G}_0$ functions, and not in the factors such as $k-p_1$, $k-p_2$ in the nominator. In the 1D case a general structure of the contribution $A(k)$ of any particular diagram of the $n$th order to $\Sigma(k)$ is
\begin{equation}
\label{gen-1D}
A(k)\propto k\int\prod\limits_{j=1}^n dp_j p_j^2 C(p_j)\prod\limits_{m=1}^{2n-1} \widetilde{G}_0\left(k-\sum\limits_{l_m} p_{l_m}\right),
\end{equation}
where every set of $l_m$ is a subset of $(1,..,n)$ and depends on the structure of the diagram. Calculating the corresponding derivatives in \eq{delta_vD0} and taking into account \eq{deriv_1D}, it is easy to see that the mER is valid, in fact, for any individual diagram.

At the same time, for the 1D case there is an exact expression for the average stationary velocity $V$ of the particle
\begin{equation}
V=\frac{D_0}{\int\limits_0^\infty dx\exp\left\{-\gamma x+g\left[1-C(x)\right]\right\}},\hskip10pt \gamma=v/D_0,
\label{1D_exact_a}
\end{equation}
which is equivalent to the full summation of the PT series for $V$ and demonstrates no singularities for any reasonable real-space correlation function $C(x)$ (i.e., when $C(0)=1$, $|C(x)| \le 1$ for $x >0$, and $C(x)\to 0$ for $x\to\infty$) \cite{Dunlap:542}. Obviously, the corresponding derivative $\partial V/\partial v$ is not singular as well. Hence, the equality between corresponding contributions to $D$ and $\partial V/\partial v$ for every diagram leads to the validity of the full mER (\ref{mER-2}) for 1D case. If needed, we may assume the proper regularization for $p\to\infty$ in every PT order, it does not affect the equality between corresponding contributions to $D/D_0$ and $\partial V/\partial v$, and the subsequent removal of regularization again leads to the desired mER.

We see that the diagrammatic technique gives a new proof of the validity of the mER, in addition to the original derivation \cite{Parris:2803}. This new proof is valid for any Gaussian random landscape and significantly extends the area of validity of the mER. Our derivation clearly shows that the mER is exclusively 1D phenomenon as it holds because of a very specific symmetry of the diagrams, where every scalar product of vectors is equivalent to a trivial multiplication of real numbers. In multidimensional case the only possibility is to to derive a series of relations between transport coefficients explicitly taking expansion into the powers of the effective charge in a manner close to Ref. \onlinecite{Hope:2377}.

\section{Behavior of $\delta V$ and $\delta D$ for $v\to 0$}

Now let us consider the behavior of $\delta D$ and $\delta V$ for small $v$. In this section we will restrict our approach to the first order PT, so we drop the corresponding index. We consider here only the isotropic random medium with spherically symmetric correlation function $C(\vec{p})=C(p)$, and the function dependence of $\delta D$ and $\delta V$ on $v$ is governed by the correlation function $C(p)$. For $v\to 0$ the most important is the long range behavior of $C(r)$ and, therefore, behavior of $C(p)$ for $p\to 0$. It is easy to show that all variety of reasonable correlation functions (we assume that $C(r)$ is a monotonously decreasing function of $r$) falls in three different classes. For example, if $C(r)\propto 1/r^{\alpha}$  for $r\to \infty$,  then for $p\to 0$
\begin{equation}
C(p)\propto
\begin{cases}
 1/p^{d-\alpha}, \hskip5pt \alpha < d,\\
 \ln(1/p), \hskip5pt \alpha = d,\\
 {\rm const}, \hskip5pt \alpha > d.
\end{cases}
\label{C(p)}
\end{equation}
Correlation functions with more faster decay (e.g., exponential or Gaussian) fall in the same class as the power law correlations with $\alpha > d$, i.e. $C(p)\propto$ const for $p\to 0$.

We should emphasize that in our consideration we exclude very long range correlations where $C(p)$ demonstrates even stronger divergence for $p\to 0$ leading to the anomalous diffusion \cite{Kravtsov:703,Kravtsov:203}. Such random energy landscapes are not expected to appear in amorphous semiconductors. Probably, the correlation of the dipolar type $C(r)\propto 1/r$ demonstrates the slowest possible decay in such materials \cite{Dunlap:542,Novikov:14573}.

Let us consider in detail the correction for $\delta V$, and then just summarize briefly the analogous results for $\delta D_{\bot}$ and $\delta D_{||}$. Let us start with the 1D case.

\subsection{1D case}

\begin{equation}
\delta V=i\frac{g D_0^2}{2\pi}\int\limits_{-\infty}^{\infty} dp C(p)\frac{p^2}{D_0p-iv}=
i\frac{g D_0}{2\pi}\int\limits_{-\infty}^{\infty} dp C(p)\left(p+i\gamma-\frac{\gamma^2}{p-i\gamma}\right)
\label{delta_v1_1D}
\end{equation}
here $\gamma=v/D_0$ and we assume that $C(p)$ is an even function of $p$. Finally
\begin{equation}
\delta V=-g v+\frac{g v^2}{D_0} \int\limits_0^\infty dx C(x) e^{-\gamma x}.
\label{delta_v1_1D_2}
\end{equation}

Hence, if the integral $\int\limits_0^\infty dx C(x)$ converges, then the leading correction to the first term in \eq{delta_v1_1D_2} is $\propto v^2$. If the integral diverges (for example, this is the case for the dipolar glass model with $C(x)\propto 1/x$), the correction is different. If $C(x)\propto 1/x^\alpha$ and $\alpha\le 1$, then the integral in \eq{delta_v1_1D_2} is effectively cut off at $x_c=1/\gamma$ and it is proportional to $\ln (1/\gamma)$ for $\alpha=1$ and $1/\gamma^{\hskip1pt 1-\alpha}$ for $\alpha < 1$. Diffusivity $\delta D$ may be obtained from $\delta V$ using the mER.

At the same time, for the 1D case we may calculate the asymptotics of $V$ at $v\to 0$ for the exact relation \eq{1D_exact_a}. The asymptotics is formed at large $x$ where $C(x)\to 0$, so
\begin{equation}
\frac{D_0}{V} \approx e^{g}\int\limits_0^\infty dx e^{-\gamma x}\left[1-g C(x)\right]=ve^{g}\left[1+\frac{gv}{D_0}\int\limits_0^\infty dx e^{-\gamma x}C(x)\right].
\label{v_exact-0}
\end{equation}
We see that behavior for $\gamma\to 0$ in \eq{delta_v1_1D_2} and \eq{v_exact-0} is the same, the only difference is that \eq{delta_v1_1D_2} gives the expansion of \eq{v_exact-0} in $g$. As we already noted in the Introduction, we may expect that this very behavior remains intact in higher dimensions. In addition, the very structure of the 1D result for $v\to 0$, i.e. the possibility to use expansion in $gC(x)$  hints for the importance of the regime of effectively small $g$ for the formation of the functional type of the dependence $V(v)$ and $D(v)$ for low $v$ and, thus, for the possibility to use the PT for the evaluation of this dependence.

\subsection{2D and 3D cases}

Isolating the maximal power of $p$ in integral (\ref{delta_v10}),   it is easy to see that $\delta V$ could be written as
\begin{eqnarray}
\vec{v}\cdot\delta \vec{V}= -\frac{g v^2\Omega_d}{(2\pi)^{d}}\int\limits_0^\infty dp p^{d-1}C(p)\left[\frac{1}{d}+M_v(p/\gamma )\right]= -g v^2\left(\frac{1}{d}+\Delta\right),\\
\Delta=\frac{\Omega_d}{(2\pi)^{d}}\int\limits_0^\infty dp p^{d-1}C(p)M_v(p/\gamma ).\nonumber
\label{DeltaV}
\end{eqnarray}
Here we performed the integration in \eq{delta_v10} over angles for the isotropic correlation function, $\Omega_d=2\pi^{d/2}/\Gamma(d/2)$ is the area of the d-dimensional sphere with unit radius and kernel $M_v(x)\to -C_d/x^2$ for $x\to\infty$, while $M_v(0)=-1/d$. Separation of the term $1/d$ in \eq{DeltaV} is motivated by vanishing of the kernel $M_v(x)$ for $x\to\infty$. For $\delta D_{||}$ and $\delta D_{\bot}$ the results which may be easily obtained by the corresponding integration of \eq{delta_D_||_20} and \eq{delta_D_bot0} have the same structure apart from the trivial replacement $v^2\Rightarrow D_0$ and, of course,  constants $C_d > 0$ are different for $\delta V$, $\delta D_{||}$, and $\delta D_{\bot}$. We see that $\Delta(v)$ provides the estimation for the mobility field dependence because $\delta\mu\propto \vec{v}\cdot\delta\vec{V}/v^2$.

For 2D case
\begin{equation}
M_v(x)=-\frac{1}{2}+x^2\left(1-\frac{x}{\sqrt{x^2+1}}\right), \hskip10pt C^v_2=\frac{3}{8},
\label{Mv-2D}
\end{equation}
and for the 3D case
\begin{equation}
M_v(x)=-\frac{1}{3}+x^2\left(1-x\arcsin\frac{1}{\sqrt{x^2+1}}\right),
\hskip10pt C^v_3=\frac{1}{5}.
\label{Mv-2D}
\end{equation}

We may obtain very rough estimation of $\Delta$ subdividing the integral over $p$ in two regions: from 0 to $\gamma $ and from $\gamma $ to $p_c$ ($p_c\simeq 1/l$ is the effective cut-off for some microscopic length scale $l$, e.g. intermolecular distance). In the first region we set $M(x)\approx M(0)$, and in the second one $M(x)\approx -C_d/x^2$. In both cases we may use for the correlation function $C(p)$ the asymptotics of small $p$ from \eq{C(p)}. Then we get
\begin{equation}
\Delta=\Delta_1+\Delta_2\simeq-\frac{\Omega_d}{(2\pi)^{d}}\left[\frac{1}{d}\int\limits_0^{\gamma } dp p^{d-1}C(p)+C_d^v\gamma^2\int\limits_{\gamma}^{p_c} dp p^{d-3}C(p)\right].
\label{Del-common}
\end{equation}
Hence, for the short range correlations with $C(p)\approx C(0)$ we have (keeping only the leading terms for $\gamma\to 0$)
\begin{eqnarray}
& \Delta_1 \simeq -\frac{\Omega_d}{(2\pi)^{d}d^2}C(0)\gamma^d,\\
& \Delta_2 \simeq -\frac{\Omega_d}{(2\pi)^{d}}C(0)C_d^v
\begin{cases}
 \gamma^2\ln(p_c/\gamma), \hskip5pt d=2,\\
 \gamma^2 p_c, \hskip5pt d=3,
\end{cases}
\label{Del-SR}
\end{eqnarray}
for the marginal case $C(p)\approx A\ln(p_c/p)$ in \eq{C(p)}
\begin{eqnarray}
& \Delta_1 \simeq -\frac{\Omega_d}{(2\pi)^{d}d^2}A\gamma^d\ln(p_c/\gamma),\\
& \Delta_2 \simeq -\frac{\Omega_d}{(2\pi)^{d}}AC_d^v \gamma^2
\begin{cases}
 \frac{1}{2}\left[\ln(p_c/\gamma)\right]^2, \hskip5pt d=2,\\
 p_c, \hskip5pt d=3,
\end{cases}
\label{Del-MC}
\end{eqnarray}
and for the long range correlation $C(p)\approx A/p^{d-\alpha}$
\begin{eqnarray}
& \Delta_1 \simeq -\frac{\Omega_d}{(2\pi)^{d}d\alpha}A\gamma^\alpha,\\
& \Delta_2 \simeq -\frac{\Omega_d}{(2\pi)^{d}}AC_d^v \gamma^2
\begin{cases}
 \frac{1}{2-\alpha}(p_c/\gamma)^{2-\alpha}, \hskip5pt \alpha < 2,\\
 p_c\ln(p_c/\gamma), \hskip5pt \alpha=2,\\
 \frac{1}{\alpha-2}, \hskip5pt \alpha > 2.
\end{cases}
\label{Del-LR}
\end{eqnarray}

Analogous results for $\delta D_{\bot}$ and $\delta D_{||}$ are
\begin{equation}
M^\bot_{2D}(x)=-\frac{x}{\sqrt{x^2+1}}\left(1+
3x^2\right)+3x^2-\frac{1}{2}, \hskip10pt C_2^{\bot}=5/8,
\label{MD_2d2-0}
\end{equation}
\begin{equation}
M^{||}_{2D}(x)=\frac{x}{\sqrt{x^2+1}}\left(1+
3x^2+\frac{x^2}{x^2+1}\right)-
3x^2-\frac{1}{2}, \hskip10pt C_2^{||}=7/8,
\label{delta_D1_2D-0}
\end{equation}
\begin{equation}
M^{||}_{3D}(x)=-5x^2+\frac{x^2}{x^2+1}+x(5x^2+1)\arcsin\frac{1}{\sqrt{x^2+1}}-\frac{1}{3},
\hskip10pt C_3^{||}=1/3,
\label{M_3D-1}
\end{equation}
\begin{equation}
M_D^\bot(x)=\frac{5}{2}x^2-\frac{x}{2}(5x^2+1)\arcsin\frac{1}{\sqrt{x^2+1}}-\frac{1}{3},
\hskip10pt C_3^{\bot}=1/3.
\label{M_3D-0}
\end{equation}
Hence, the corrections for the field dependences of $\delta D_{\bot}$ and $\delta D_{||}$ could be obtained from the corresponding corrections for $\vec{v}\cdot\vec{V}$ by the trivial replacement of the constant $C$ and $v^2\Rightarrow D_0$.

\begin{figure}[tbp]
\includegraphics[width=3in]{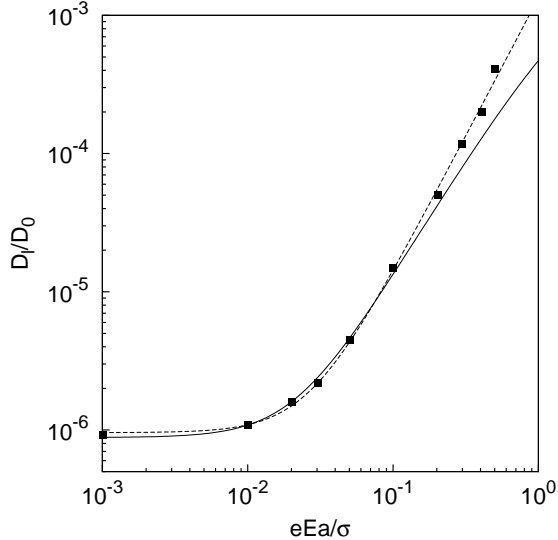}
\caption{Fit of data from Ref. \onlinecite{Nenashev:115204} (filled squares) for the dependence $D/D_0=A+B(eEa/\sigma)^2\ln(E/E_0)$ (solid line), $eE_0a/\sigma\approx 4$, $A\approx 8.8\times 10^{-7}$, $B\approx -7.8\times 10^{-4}$. Broken line demonstrates the best fit of the data for the dependence $D/D_0=A+B(eEa/\sigma)^2$}  \label{fit2D}
\end{figure}

We see that the behavior for the 3D GDM agrees well with the result of the computer simulation \cite{Nenashev:115204}, but the 2D case does differ and contains an additional logarithmic factor. It is rather difficult to catch such slowly varying factor in addition to the major contribution $\propto \gamma^2$ while analyzing the simulation data, especially taking into account the limited accuracy of the simulation data. For this reason the logarithmic factor has not been found in Ref. \onlinecite{Nenashev:115204}. To support this statements we provide the fit of the data for 2D longitudinal diffusivity borrowed from Ref. \onlinecite{Nenashev:115204} using \eq{Del-SR} (see Fig. \ref{fit2D}). We do not pretend to provide a proper description of the data from Ref. \onlinecite{Nenashev:115204} with our formula, this is clearly impossible due to the limitation $(\sigma/kT)^2\ll 1$ for our approach. In Fig \ref{fit2D} we just demonstrate the difficulty to distinguish the dependences $\propto \textrm{const}+E^2\ln E$ and $\propto \textrm{const}+E^2$ for $E\to 0$. Indeed, a significant difference between both depenedences arises only for fields where parameter $eaE/\sigma$ becomes comparable to 1.

\section{Experimental evidence for the validity or invalidity of the Einstein relation in amorphous organic semiconductors}

The Gaussian DOS is considered as the most appropriate model for amorphous organic materials \cite{Bassler:15}. Validity of the ER in amorphous organic semiconductors demonstrating hopping charge transport is still a controversial question. There are reliable theoretical results showing that the ER cannot hold for the materials having the Gaussian DOS and demonstrating the non-linear average velocity dependence on $E$ or having a non-negligible concentration of charge carriers \cite{Parris:2803,Roichman:1948}. Invalidity of the Einstein relation in amorphous materials is supported also by computer simulation \cite{Novikov:391}.

For the experimental test of the validity of the ER the most suitable is the so-called quasi-equilibrium transport regime where all initial carrier relaxation is over and carrier velocity (averaged over short time intervals) becomes constant. One of the widely used technique for a direct measurement of the  charge carrier velocity is the time-of-flight experiment \cite{Bassler:15}. In this experiment the quasi-equilibrium regime manifests itself by the development of the plateau of the current transient indicating the constant average carrier velocity.

Recent paper by Wetzelaer et al. (Ref. \onlinecite{Wetzelaer:66605}) states that in quasi-equilibrium regime the simple ER perfectly holds if we remove the influence of deep traps. They made the conclusion using rather indirect experimental evidence on the luminance of the organic light-emitting diodes. Very probably, the approximate validity of the simple ER is due to the low applied electric field, where the ER indeed holds (see Fig. 3 in Ref. \onlinecite{Wetzelaer:66605}, where $E < 7\times 10^4$ V/cm which is rather weak field). We should note also that for some materials studied in Refs. \onlinecite{Wetzelaer:66605} and \onlinecite{Nicolai:172107} (for example, for  poly(9,9-dioctylfluorene)) the reported mobility differs by approximately two orders of magnitude with the previously reported values \cite{Redecker:1565,Kreouzis:235201}. This difference hints to the rather unusual structure of the thin transport layers used in light emitting diodes (may be, the structure of the layer is not spatially uniform), which provides an additional complicating factor.

\begin{figure}[tbp]
\includegraphics[width=3in]{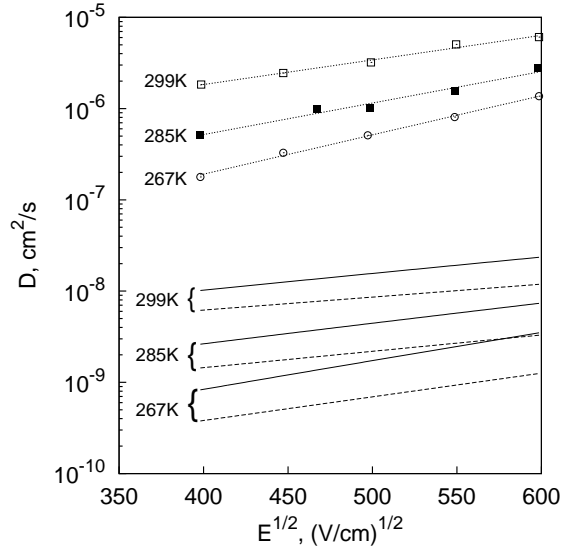}
\caption{Field dependence of the diffusivity in molecularly doped polymer. Points are borrowed from Ref. \onlinecite{Hirao:1787}, the temperature is indicated at the left. Dotted lines are shown as a guide for an eye. Solid lines show the diffusivity calculated from the experimental mobility values assuming $\ln\mu=A+B\sqrt{E}$ and using the mER \eq{mER}. Broken lines show the diffusivity calculated using the simple ER \eq{ER}.} \label{Hirao}
\end{figure}

We believe that the papers of the Nishizawa group provide much more clear direct evidence on the validity of the ER \cite{Hirao:1787,Hirao:4755,Hirao:2904,Hirao:12991,Nishizawa:L250}. They extracted $\mu$ and $D$ by fitting the experimental time-of-flight transients in various molecularly doped polymers with the solution of classic diffusion-drift equation. Typically, the quality of  fits is rather good (see Refs. \onlinecite{Hirao:1787,Hirao:4755,Hirao:12991}). Moreover,  obtained transport parameters $\mu$ and $D$ show no dependence on the thickness of transport layers, thus indicating a well-established quasi-equilibrium transport regime \cite{Hirao:4755}. At the same time, the difference between fitted $D$ and calculated using the simple or modified ER is about two orders of magnitude (see Fig. \ref{Hirao}). Such huge difference strongly support the idea of the invalidity of any variant of Einstein relation for 3D charge transport in amorphous materials with the Gaussian DOS.

Unfortunately, the direct comparison of our results for the behavior of $V$ and $D$ in the limit of weak driving force with the experimental data on charge carrier transport cannot be done due to the total lack of the reliable data for very weak field region.

\section{Conclusion}

We considered the diffusive motion of a particle in the random spatially correlated energy landscape having the Gaussian DOS. For such system the average particle velocity in the quasi-equilibrium regime is a nonlinear function of the driving force and the simple Einstein relation is certainly not valid. Using the perturbation theory we found that the modified Einstein relation \cite{Parris:2803} is an exclusively 1D property and does not hold for higher dimensions $d > 1$. For this reason a usual estimation of the diffusivity from the mobility could be approximately valid only for a low force region because the simple Einstein relation which is certainly valid at zero driving force serves as a kind of anchor point here.

We provide also a new proof of the mER for 1D case which is completely different from the previous one \cite{Parris:5295,Parris:2803}. This new proof extends the validity of the mER to arbitrary Gaussian random landscape and  does not depend on the assumption of the particular type of  correlation function, thus covering a more wider variety of possible random landscapes.

We obtained also the leading corrections for the average velocity and diffusivity in the limit of weak driving force and demonstrated how such corrections depend on the dimensionality and correlated properties of the random landscape. For the short range correlation we obtain results which agree well with the corresponding dependences for the lattice model \cite{Nenashev:115203,Nenashev:115204}. At the same time, the results show that the case $d=1$ is not exceptional one and the functional form of the corrections vary in some regular way with the variation of $d$.

\section*{Acknowledgements}
Financial support from the Program of Basic Research of the National Research University Higher School of Economics is gratefully acknowledged.


\appendix
\section{Diagrammatic technique for the particle diffusion in random  medium}
\label{appA}

We consider the diffusion of the charged particle in the random environment having the spatially correlated Gaussian DOS. For the particular realization of the random potential $U(\vec{x})$ the particle Green function $G_U(\vec{x},t)$ obeys \eq{G(x,t)}. If we consider the Laplace transform according to $t$ and Fourier transform for $\vec{x}$, then the corresponding equation becomes
\begin{equation}
G_U(\vec{k},s)=G_0(\vec{k},s)\left[1-\frac{\beta D_0}{(2\pi)^d}\int d\vec{p} G_U(\vec{k}-\vec{p},s)U(\vec{p}) \left(\vec{k}\cdot\vec{p}\right)\right],\hskip10pt G_0^{-1}(\vec{k},s)=s+D_0k^2+i\vec{v}\cdot\vec{k},
\label{G(k)}
\end{equation}
here $G_0(\vec{k},s)$ is the Green function for the zero disorder. In future we are going to consider the stationary case $s=0$ only, and use the simplified notation $G_U(\vec{k},0)=G_U(\vec{k})$. Considering the PT expansion of \eq{G(k)} and making the average over disorder, we may write down the diagram expansion for the averaged over disorder `Green function $G(\vec{k})=\left<G_U(\vec{k})\right>$. Details of the diagram technique may be found in excellent Bouchaud and Georges review \cite{Bouchaud:127}. The trivial difference between out and their notations is that they used the random force $\vec{F}=-\nabla U$ instead of $U$.

\begin{figure}
\includegraphics[width=1.7in]{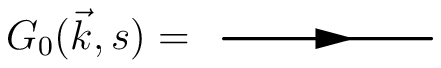}
\begin{center}a)\end{center}

\vskip10pt

\includegraphics[width=1.7in]{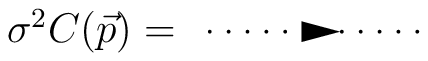}
\begin{center}b)\end{center}

\vskip10pt

\includegraphics[width=2.4in]{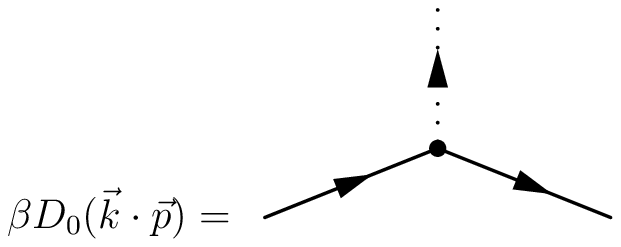}
\begin{center}c)\end{center}

\caption{a) Bare Green function $G_0(\vec{k},s)$.
b) Correlation function of the Gaussian random field.
c) Interaction vertex $\beta D_0 (\vec{k}\cdot \vec{p})$.}
\label{FD-blocks}
\end{figure}

The basic building blocks of a diagram are shown in Fig. \ref{FD-blocks}. For every inner moment $\vec{p}$ there is an integration $\frac{1}{(2\pi)^d}\int d\vec{p}$, and dotted line with $\vec{p}$ going \textit{into} the vertex provide $-\vec{p}$ for the vertex weight because of the momentum conservation
\begin{equation}
\left<U(\vec{k}_1)U(\vec{k}_2)\right>=(2\pi)^d \sigma^2 \delta(\vec{k}_1+\vec{k}_2)C(\vec{k}_1),
\label{Gaussian U(x)2}
\end{equation}
where $C(\vec{k})$ is the Fourier transform of the spatial correlation function $C(\vec{x})=\left<U(\vec{x})U(0)\right>/\sigma^2$.
We assume that the integrals converge at $p\to\infty$, as it is the case for fast decaying $C(p)$. Moreover, keeping in mind the possible application of the theory to the diffusion of particles in amorphous material we should expect an inevitable cut-off at $p\simeq 1/l$, where $l$ is some typical atomic or molecular scale.

\begin{figure}[h]
\includegraphics[width=1.5in]{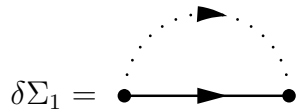}
\caption{First order contribution to $\Sigma(\vec{k})$.}
\label{fSig1}
\end{figure}

Here we briefly show only the expansion for the self-energy $\Sigma(\vec{k})=G_0^{-1}(\vec{k})-G^{-1}(\vec{k})$ where only the strongly connected diagrams should be taking into account (the diagrams which cannot be disconnected by cutting a $G_0$ line). The first order contribution to $\Sigma(\vec{k})$ is (see Fig. \ref{fSig1})

\vskip20pt

\begin{figure}[h]
\includegraphics[width=5.5in]{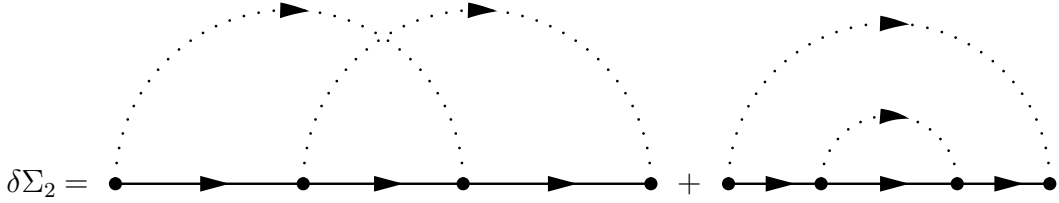}
\caption{Second order contribution to $\Sigma(\vec{k})$.}
\label{fSig2}
\end{figure}

\begin{equation}\label{fSig1e}
\delta\Sigma_1=-\frac{g D_0^2}{(2\pi)^d}\int d\vec{p}C(\vec{p})G_0(\vec{k}-\vec{p})(\vec{k}\cdot \vec{p})\left[(\vec{k}-\vec{p})\cdot \vec{p}\right],\hskip10pt g=(\sigma\beta)^2,
\end{equation}

\noindent
and the second order one is (see Fig. \ref{fSig2})


\begin{eqnarray}
\delta\Sigma_2&=\frac{g^2 D_0^4}{(2\pi)^{2d}}\int d\vec{p_1}d\vec{p_2}C(\vec{p}_1)C(\vec{p}_2)G_0(\vec{k}-\vec{p}_1)
G_0(\vec{k}-\vec{p}_1-\vec{p}_2)(\vec{k}\cdot \vec{p}_1)\left[(\vec{k}-\vec{p}_1)\cdot \vec{p}_2\right]\times \nonumber \\
&\left\{\left[(\vec{k}-\vec{p}_1-\vec{p}_2)\cdot\vec{p}_1\right]R(\vec{k},\vec{p}_2)
+
\left[(\vec{k}-\vec{p}_1-\vec{p}_2)\cdot\vec{p}_2\right]R(\vec{k},\vec{p}_1)
\right\}, \\
&R(\vec{k},\vec{p})=(\vec{k}-\vec{p})\cdot \vec{p}G_0(\vec{k}-\vec{p}).\nonumber
\label{2nd_order}
\end{eqnarray}
We mostly use the first order approximation for $\Sigma(\vec{k})$, and the second order one is used only for the test of the validity of the modified Einstein relation.

\section{Second order PT approximation}
\label{appB}

Calculation of the second order PT corrections to average velocity and diffusivity is rather straightforward but produces  complicated expressions, so we write down here only the relevant ingredients for the test of the validity of the mER \eq{nD-mER-20}
\begin{eqnarray}
&\sum\limits_a\frac{\partial \delta V^{(2)}_a}{\partial v_a}=\frac{g^2 D_0^4}{(2\pi)^{2d}}\int d\vec{p_1}d\vec{p_2}C(\vec{p}_1)C(\vec{p}_2)\times\nonumber\\
&\left\{\left[\vec{F}(\vec{p}_1)\cdot \vec{F}(\vec{p}_2)\right]\left[\vec{p}_1\cdot \vec{F}(\vec{p}_1+\vec{p}_2)\right]\vec{p}_2^{\hskip2pt 2}\left(\vec{p}_1\cdot \left[\vec{F}(\vec{p}_1)+\vec{F}(\vec{p}_1+\vec{p}_2)+\vec{F}(\vec{p}_2)\right]\right)+\right.\ \\
&\left.+\left[\vec{p}_1\cdot\vec{F}(\vec{p}_1)\right]\left[\vec{p}_2\cdot\vec{F}(\vec{p}_1)\right]
\left[\vec{p}_2\cdot\vec{F}(\vec{p}_1+\vec{p}_2)\right]
\left(\vec{p}_1\cdot\left[2\vec{F}(\vec{p}_1)+\vec{F}(\vec{p}_1+\vec{p}_2)\right]\right)\right\},\nonumber\\
&\vec{F}(\vec{p})=\vec{p}G_0(-\vec{p}).\nonumber
\label{2nd_order_Vaa}
\end{eqnarray}

\begin{eqnarray}
&\frac{1}{D_0}\sum\limits_a \delta D^{(2)}_{aa}=\frac{g^2 D_0^4}{(2\pi)^{2d}D_0}\int d\vec{p_1}d\vec{p_2}C(\vec{p}_1)C(\vec{p}_2)\times\nonumber\\
&\left\{\vec{p}_1\cdot \left[\vec{K}(\vec{p}_1)+\vec{K}(\vec{p}_1+\vec{p}_2)+\vec{K}(\vec{p}_2)\right]
\left[\vec{p}_2\cdot \vec{F}(\vec{p}_1)\right]
\left[\vec{p}_1\cdot \vec{F}(\vec{p}_1+\vec{p}_2)\right]
\left[\vec{p}_2\cdot \vec{F}(\vec{p}_2)\right]+\right.\nonumber\\
&+\left.\vec{p}_1\cdot \left[2\vec{K}(\vec{p}_1)+\vec{K}(\vec{p}_1+\vec{p}_2)\right]
\left[\vec{p}_2\cdot \vec{F}(\vec{p}_1)\right]
\left[\vec{p}_2\cdot \vec{F}(\vec{p}_1+\vec{p}_2)\right]
\left[\vec{p}_1\cdot \vec{F}(\vec{p}_1)\right]-\right.\nonumber\\
&-\left.\left[\vec{p}_2\cdot \vec{F}(\vec{p}_1)\right]
\left[\vec{p}_1\cdot \vec{F}(\vec{p}_1+\vec{p}_2)\right]
\left[\left(\vec{p}_1+\vec{p}_2\right)\cdot \vec{F}(\vec{p}_2)\right]-\right.\\
&-\left.\left[\vec{p}_1\cdot \vec{F}(\vec{p}_1)\right]
\left[\vec{p}_2\cdot \vec{F}(\vec{p}_2)\right]
\left[\vec{p}_1\cdot \vec{p}_2 G_0(-\vec{p}_1-\vec{p}_2)\right]-\right.\nonumber\\
&\left.-2\left[\vec{p}_1\cdot \vec{F}(\vec{p}_1)\right]
\left[\vec{p}_2\cdot \vec{F}(\vec{p}_1)\right]
\left[\vec{p}_2\cdot \vec{F}(\vec{p}_1+\vec{p}_2)\right]-
\left[\vec{p}_2\cdot \vec{F}(\vec{p}_1)\right]^2\vec{p}_1^{\hskip2pt 2}
G_0(-\vec{p}_1-\vec{p}_2)\right\},\nonumber\\
&\vec{K}(\vec{p})=G_0(-\vec{p})\left(2D_0\vec{p}-i\vec{v}\right).\nonumber
\label{2nd_order_Daa}
\end{eqnarray}
Right parts of \eq{2nd_order_Vaa} and (\ref{2nd_order_Daa}) do differ for any $d > 1$.


\end{document}